\documentclass[secnumarabic, superscriptaddress, amssymb, nobibnotes, aps, prmaterials, twocolumn, longbibliography]{revtex4-2}

\usepackage{amsmath}
\usepackage{graphicx}
\usepackage{mathtools}
\usepackage{xcolor}
\usepackage{todonotes}
\usepackage{hyperref}
\usepackage{gensymb}
\usepackage{ulem}
\usepackage{multirow}
\usepackage{xcolor}

\begin{document}
\title{Evolution of the Saddle Point in 
Antimony Telluride Homologous Superlattices}

\author{Yi-Hsin Shen}
\affiliation{Applied Physics Program, University of Michigan, Ann Arbor, MI 48109, USA}

\author{Shane Smolenski}
\affiliation{Department of Physics, University of Michigan, Ann Arbor, MI 48109, USA}

\author{Ming Wen}
\affiliation{Department of Chemistry, University of Michigan, Ann Arbor, MI 48109, USA}

\author{Yimo Hou}
\affiliation{Department of Materials Science and Engineering, University of Michigan, Ann Arbor, MI 48109, USA}

\author{Eoghan Downey}
\affiliation{Department of Physics, University of Michigan, Ann Arbor, MI 48109, USA}

\author{Jakob Hammond-Renfro}
\affiliation{Department of Physics, University of Michigan, Ann Arbor, MI 48109, USA}

\author{Katharine Moncrieffe}
\affiliation{Department of Electrical Engineering, Cooper Union for the Advancement of Science and Art, New York City, NY 10003, USA}

\author{Chun Lin}
\affiliation{Stanford Synchrotron Radiation Lightsource, SLAC National Accelerator Laboratory, Menlo Park, CA 94025, USA}

\author{Makoto Hashimoto}
\affiliation{Stanford Synchrotron Radiation Lightsource, SLAC National Accelerator Laboratory, Menlo Park, CA 94025, USA}

\author{Donghui Lu}
\affiliation{Stanford Synchrotron Radiation Lightsource, SLAC National Accelerator Laboratory, Menlo Park, CA 94025, USA}

\author{Kai Sun}
\affiliation{Department of Physics, University of Michigan, Ann Arbor, MI 48109, USA}

\author{Dominika Zgid}
\affiliation{Department of Physics, University of Michigan, Ann Arbor, MI 48109, USA}
\affiliation{Department of Chemistry, University of Michigan, Ann Arbor, MI 48109, USA}
\affiliation{Institute of Theoretical Physics, University of Warsaw, 02-093 Warsaw, Poland }

\author{Emanuel Gull}
\affiliation{Department of Physics, University of Michigan, Ann Arbor, MI 48109, USA}
\affiliation{Institute of Theoretical Physics, University of Warsaw, 02-093 Warsaw, Poland }

\author{Pierre Ferdinand P. Poudeu}
\affiliation{Department of Materials Science and Engineering, University of Michigan, Ann Arbor, MI 48109, USA}

\author{Na Hyun Jo}
\altaffiliation{nhjo@umich.edu}
\affiliation{Department of Physics, University of Michigan, Ann Arbor, MI 48109, USA}

\author{Rachel S. Goldman}
\altaffiliation{rsgold@umich.edu}
\affiliation{Department of Materials Science and Engineering, University of Michigan, Ann Arbor, MI 48109, USA}
\affiliation{Department of Physics, University of Michigan, Ann Arbor, MI 48109, USA}

\date{\today}

\def\kill #1{\sout{#1}}
\def\add #1{\textcolor{blue}{#1}} 
\def\addred #1{\textcolor{red}{#1}} 
\newcommand{\tocite}[1]{$^\textbf{\textcolor{red}{CITE} #1}$ }
\newcommand{\postit}[1]{
    \begin{center}
        \fbox{
            \begin{minipage}{0.4\textwidth}
                \textit{#1}
            \end{minipage}
            }
    \end{center}
}

\begin{abstract}
Combining topological insulators with topological semimetals in the form of homologous superlattices is a promising approach for generating correlated quantum matter based upon Fermi level alignment with band extrema. For antimony telluride, a saddle point is predicted to occur at the M-point, while antimonene layering is predicted to move the M-point valence band towards the Fermi level. To date, the predicted saddle point at the M-point has not yet been demonstrated, and studies of antimony telluride homologous superlattices have been limited to one or two layers of antimonene added to antimony telluride.  Here, we present scanning tunneling spectroscopy and angle-resolved photoemission spectroscopy studies of a series of antimony telluride homologous superlattices with two to four layers of antimonene.  In addition to demonstrating the presence of a saddle point and associated van Hove singularity near the M-point, we identify the key role of Sb and Te $p_z$ orbital hybridization in driving the van Hove singularity toward the Fermi level. 
 
\end{abstract}

\maketitle

\section{Introduction}

Engineering the density of states near the Fermi level is a central strategy for accessing correlated quantum matter, with broad implications for emerging electronic and quantum technologies  ~\cite{stormer1999fractional, bistritzer2011moire}. One particularly powerful route involves band extrema such as saddle points, which generate van Hove singularities (VHSs) with divergent density of states ~\cite{van1953occurrence, tan2021charge, hlubina1997ferromagnetism, hirsch1986enhanced}. To date, approaches for bringing the Fermi level into alignment with a VHS have relied mainly upon gating  ~\cite{kim2023electric}, chemical doping ~\cite{rosenzweig2019tuning}, or heterostructure engineering ~\cite{mori2019controlling}. Here, we demonstrate an alternative pathway - combining topological insulators with topological semimetals in the form of homologous superlattices.


For the topological semiconductor, Sb\textsubscript{2}Te\textsubscript{3}, a saddle point is predicted to occur at the M-point, albeit several 100s of meV below the Fermi level~\cite{osti_1188507, mohelsky2024electronic}. Meanwhile, layering of the topological semi-metal antimonene (Sb\textsubscript{2}) is predicted to move the M-point valence band towards the Fermi level ~\cite{aresFewlayerAntimoneneElectrical2021}.  Furthermore, incorporation of layers of antimonene (Sb\textsubscript{2}) into Sb\textsubscript{2}Te\textsubscript{3} reduces the Sb\textsubscript{2}Te\textsubscript{3} band gap, driving a transition from semiconducting to semi-metallic behavior ~\cite{johannsen2015engineering, khalil2017electronic}. Therefore, we hypothesize that antimony telluride homologous superlattices ~\cite{poudeu2005design, kifune2005extremely, shelimova2000homologous} with multiple antimonene layers will facilitate alignment of the VHS and the Fermi level without the need for gating or doping. 

To date, reports on the addition of one and two layers of antimonene to Sb\textsubscript{2}Te\textsubscript{3}, i.e. (Sb\textsubscript{2}Te\textsubscript{3})\textsubscript{1}(Sb\textsubscript{2})\textsubscript{1} and (Sb\textsubscript{2}Te\textsubscript{3})\textsubscript{1}(Sb\textsubscript{2})\textsubscript{2} antimony telluride homologous superlattices, have revealed band gap closing with retained topological surface states ~\cite{cecchi2019interplay, johannsen2015engineering}. However, the saddle point has not yet been confirmed experimentally and the influence of more than two antimonene layers on the electronic structure remains unknown. 
Using scanning tunneling spectroscopy (STS) and angle-resolved photoemission spectroscopy (ARPES), in conjunction with with sc$GW$ calculations, we provide evidence for a saddle point and associated van Hove singularity near the M-point.  
We identify the key role of Sb and Te $p_z$ orbital hybridization in generating the saddle point and shifting it to higher energies with increasing number of antimonene layers.  These new insights into antimonene-induced band structure evolution in antimony telluride homologous superlattices 
provide a crucial step for generating correlated quantum matter in a wide range of systems.


\section{Methods}
For these studies, a series of single crystals were prepared using solution (self-flux) growth and solid-state synthesis. For solution-growth, an 80-20 stoichiometric mixture of 99.999\% pure Sb shots (ThermoScientific) and 99.999+\% lump Te (ThermoScientific) 
were placed into a fritted alumina crucible~\cite{CANFIELD2016,SLADE2022} and sealed in a quartz ampoule under partial pressure of argon. The ampoule was heated to 700\degree C over 5 hours, held at this temperature for 5 hours, and then cooled to 460\degree C over 20 hours, 
at which point the excess solution was separated from the Sb\textsubscript{2}Te\textsubscript{3} single crystals using a centrifuge. 
To prepare antimony telluride homologous superlattices with two to four layers of antimonene, i.e.
(Sb\textsubscript{2}Te\textsubscript{3})\textsubscript{1}(Sb\textsubscript{2})\textsubscript{n} with n = 2, 3, 4 (Fig.~\ref{fig:structure}(a)), stoichiometric mixtures of 99.5\% Sb (Alfa Aesar, 200 mesh) and 99.999\% Te (ThermoScientific, 100 mesh) were 
pre-ground 
in an argon-filled glove box 
and loaded into evacuated quartz tubes. 
Following five homogenization cycles consisting of annealing  
at 720\degree C (100-150 minutes), followed by furnace cooling to 550\degree C, 
the samples were held at 550\degree C for 2 months, followed by a 3-hour ramp down to room temperature.

The crystal structures and long-period stacking sequences of the homologous superlattices were determined using an analysis of powder x-ray diffraction (XRD) patterns generated using Cu K$\alpha$ radiation.  In particular, $\theta$--$2\theta$ scans were collected in a Rigaku Smart Lab diffractometer operating at 40 kV/44 mA using a wide-open detector.


In preparation for spectroscopic studies in ultra-high vacuum (UHV), aluminum posts were mounted on the ingot surfaces using silver paste. Within each UHV chamber, the aluminum post was cleaved off using a wobble stick, leaving freshly cleaved surfaces for scanning tunneling microscopy/spectroscopy (STM/STS), ARPES, and x-ray photoelectron spectroscopy (XPS). In all cases, the base pressures were $\leq$ {3} $\times$ 10$^{-11}$\,Torr. 

STM and STS were performed 
at room temperature and 77\,K using commercially-available PtIr tips, cleaned in-situ by electron bombardment. STS was performed in constant-height mode (i.e. fixed tip-sample separation), with a continuous linear voltage ramp applied to the sample and an AC voltage supplied by a lock-in amplifier.~\cite{liu2025probing} The modulation in applied voltage produces a modulation of the tunneling current; the resulting differential conductance, d$I$/d$V$, and the second-harmonic of the tunneling current, d$^2I$/d$V^2$, as a function of bias voltage ~\cite{miyamoto1989tunneling} are outputs of the lock-in amplifier.

ARPES and XPS were performed at Beamline 5-2 of the Stanford Synchrotron Radiation Lightsource at the SLAC National Accelerator Laboratory using a Scienta DFS30 Electron Analyzer. 
Data were acquired at a base temperature of $\sim 7.5$\,K using linear horizontal polarized light. Fixed photon energy ARPES was performed at $h\nu\,=\,100$\,eV for core-level spectra and in the energy range $h\nu\,=\,89$\,-\,$93$\,eV for valence band measurements. The beam spot size was $\sim50$ $\times$ 10\,$\mu$m\textsuperscript{2}. 

For computational studies, (Sb$_2$Te$_3$)$_1$(Sb$_2$)$_2$ was modeled as a repeat unit consisting of a Sb$_2$Te$_3$ quintuple layer followed by two bilayers of Sb$_2$ (antimonene), as shown in ~\ref{fig:structure}(a). The band structure was computed using the self-consistent $GW$ method using a Gaussian type orbital (GTO) basis. 
A $6\times 6\times 6 $ \textit{k}-mesh sampling was used for the hexagonal (Sb$_2$Te$_3$)$_1$(Sb$_2$)$_2$ cell. 
The initial mean-field DFT calculation was performed with \texttt{pyscf}.~\cite{sunLibcintEfficientGeneral2015a,sunPySCFPythonbasedSimulations2018,sunRecentDevelopmentsPySCF2020} We chose the combination of PBE functional~\cite{Perdew96} and the gth-szv-molopt-sr basis set.~\cite{VandeVondele07} The corresponding gth-pbe pseudopotentials were also used for all atoms.~\cite{Goedecker96}
The following sc$GW$ calculation was performed on the Matsubara axis under finite-temperature (inversed temperature $\beta = 1000 $ a.u.) with the \texttt{Green/WeakCoupling} code suite.~\cite{Iskakov20,Pokhilko22,yehFullySelfconsistentFinitetemperature2022a,Green23,iskakovGreenWeakCouplingImplementation2025a}
Spectral functions and density of states (DOS) were analytically continued from the converged Matsubara Green's function with Nevanlinna analytic continuation technique.~\cite{feiNevanlinnaAnalyticalContinuation2021}

\section{Superlattice Periodicities}
\begin{figure*}[ht]
    \centering
    \includegraphics[width=\linewidth]{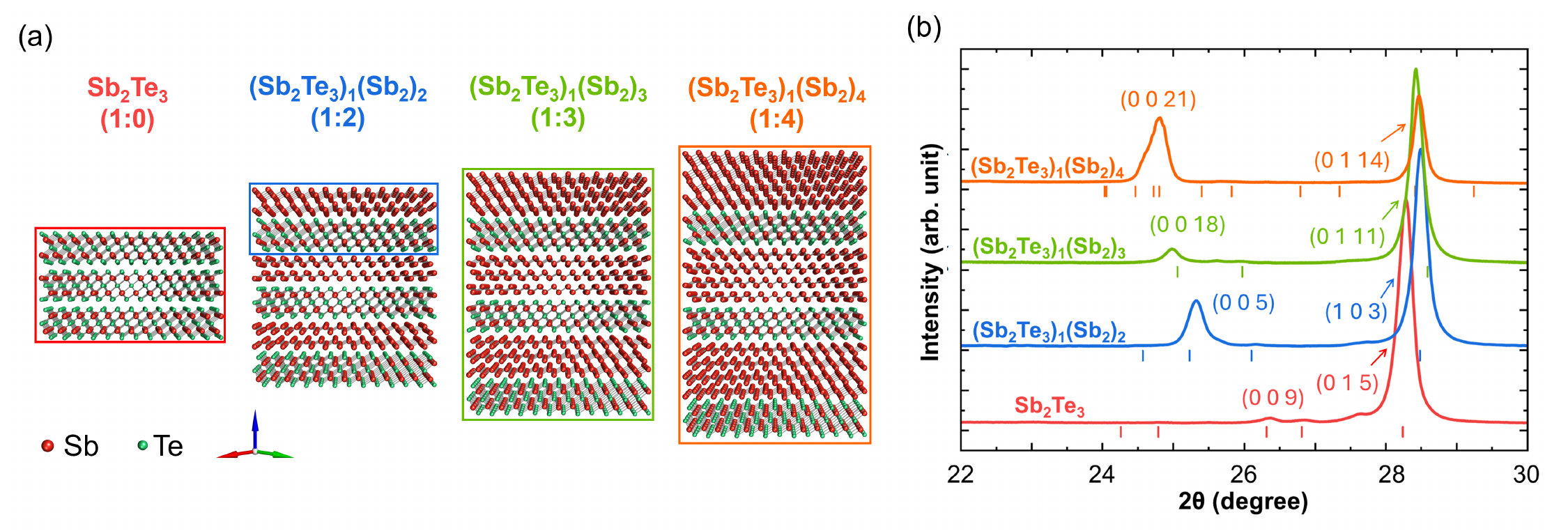}
    \caption{(a) Models for antimony telluride homologous superlattices, with Sb and Te atoms represented as red and green circles. The repeat units, illustrated with colored boxes, consists of antimony telluride (Sb\textsubscript{2}Te\textsubscript{3}), with or without two to four layers of antimonene (Sb\textsubscript{2}).
    Sb\textsubscript{2}Te\textsubscript{3} (red) and homologous superlattices containing three (green) and four (orange) layers of antimonene  
    belong to the rhombohedral $R$-3M space group with ABC stacking, while antimony telluride homologous superlattices with two layers of antimonene (blue) 
    crystallize in the trigonal $P$-3M1 space group. (b) Powder X-ray diffraction (XRD) patterns for  Sb\textsubscript{2}Te\textsubscript{3} (red) and homologous superlattices containing two (blue), three (green), and four (orange) layers of antimonene. 
    The diffraction patterns are vertically offset for clarity. Peaks are indexed according to the average-structure assignments obtained from whole-pattern analysis of the full powder XRD data.}
    \label{fig:structure}
\end{figure*}

We first identify the crystal structures of the antimony telluride superlattices based upon their long-period stacking sequences.  In Fig.~\ref{fig:structure}(b), powder XRD patterns are shown for Sb\textsubscript{2}Te\textsubscript{3} (red) and homologous superlattices containing two (blue), three (green), and four (orange) layers of antimonene, 
with representative reflections indexed in the plot. The patterns are dominated by out-of-plane $(00l)$ reflections associated with the long-period stacking sequences.
Whole-pattern indexing/refinement of the full powder XRD data indicates that antimony telluride (red) and homologous superlattices containing three (green) and four (orange) antimonene layers are consistent with a rhombohedral $R$-3M structure, whereas the homologous superlattices containing two (blue) antimonene layers are better described by the trigonal $P$-3M1 structure ~\cite{poudeu2005design, kifune2005extremely}.
Here, the symmetry labels are used as average-structure descriptions that best account for the laboratory XRD patterns,
without over-interpreting subtle stacking/order variants that may be present in these long-period layered compounds.
With increasing number of antimonene layers, the dominant out-of-plane reflections shift to lower $2\theta$ and the spacing between successive peaks decreases,
indicating a systematic evolution of the out-of-plane periodicity as additional antimonene layers are inserted.

\section{semiconductor to semimetal}
\begin{figure*}[ht]
    \centering
    \includegraphics[width=0.85\linewidth]{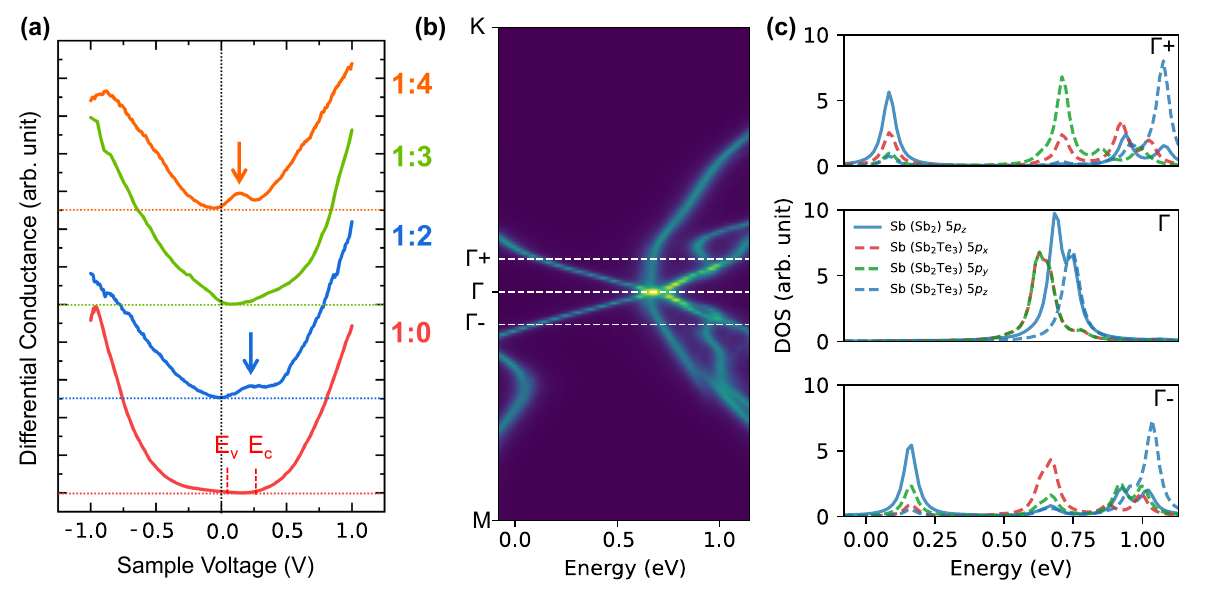}
    \caption{(a) Plots of differential conductance vs. sample bias voltage, collected at 77K, for antimony telluride homologous superlattices with 2, 3, and 4 antimonene layers, i.e. (Sb\textsubscript{2}Te\textsubscript{3})\textsubscript{1} (Sb\textsubscript{2})\textsubscript{n} (n = 2, 3, 4), in comparison with that of antimony telluride, Sb\textsubscript{2}Te\textsubscript{3}. The sample voltages correspond to the energy relative to the Fermi level, and all spectra are vertically offset for clarity.  For Sb$_{2}$Te$_{3}$, well-defined band-edges, with a band gap of $\approx$ 0.22 eV, is apparent, consistent with earlier reports \cite{orlov2019bismuth, mohelsky2024electronic}. 
    In contrast, for the antimony telluride homologous superlattices, the band gap closes \cite{johannsen2015engineering} and additional states emerge above the Fermi level, as indicated by downward pointing arrows. (b) \textit{GW}-computed electronic energy band dispersion in the vicinity of the $\Gamma$ point with white dashed lines marking points within $\pm$ 0.12 \AA\textsuperscript{-1} of the $\Gamma$-point. (c) Selected atomic orbital contributions to the computed density of states at $\Gamma$\textsubscript{+} (top), $\Gamma$ (middle), and $\Gamma$\textsubscript{--} (bottom). Only specific AOs are shown for better clarity.
    See the Supplemental Material for a more detailed DOS decomposition.} 
    \label{fig:STS}
\end{figure*}

To understand the influence of antimonene on the semiconductor to semimetal transition in antimony telluride homologous superlattices, we consider STS data in conjunction with sc$GW$ calculations. In Fig.~\ref{fig:STS}(a), the differential conductance versus sample bias voltage is shown for antimony telluride homologous superlattices with 2, 3, and 4 antimonene layers, shown in blue, green and orange, in comparison with that of antimony telluride, shown in red. The sample voltages correspond to the energy relative to the Fermi level, and all spectra are vertically offset for clarity.  For antimony telluride, well-defined band-edges, with a band gap of $\approx$ 0.22 eV, is apparent, consistent with earlier reports \cite{orlov2019bismuth, mohelsky2024electronic}. In contrast, for the antimony telluride homologous superlattices, the band gap closes \cite{johannsen2015engineering} and additional states emerge above the Fermi level, as indicated by downward pointing arrows in Fig.~\ref{fig:STS}(a). 

\begin{figure*}[ht]
    \centering
    \includegraphics[width=0.85\linewidth]{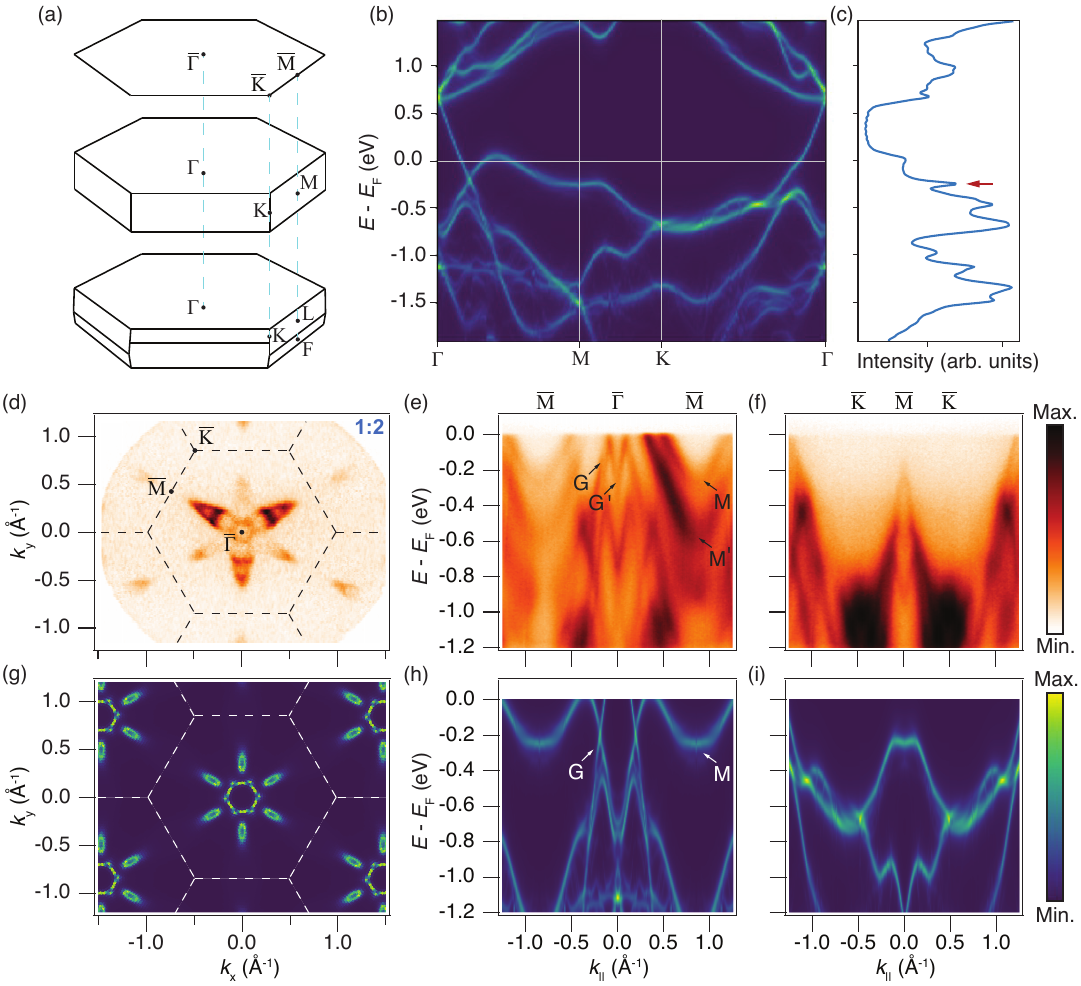}
    \caption{Electronic structure of antimony telluride homologous superlattices. (a) Surface and bulk Brillouin zones (BZ), including the trigonal surface BZ for both antimony telluride and its homologous superlattices (top), the hexagonal bulk BZ for antimony telluride homologous superlattices containing two layers of antimonene (middle), and the rhomobohedral bulk BZ for antimony telluride homologous superlattices with three and four layers of antimonene (bottom).(b) Computed electron energy band dispersion along high symmetry paths for antimony telluride homologous superlattices containing two layers of antimonene, (Sb\textsubscript{2}Te\textsubscript{3})\textsubscript{1}(Sb\textsubscript{2})\textsubscript{2}. (c) Computed density of states as a function of energy for (Sb\textsubscript{2}Te\textsubscript{3})\textsubscript{1}(Sb\textsubscript{2})\textsubscript{2}, summed along the $\Gamma$-M-K-$\Gamma$ path. The peak at $\sim -200$\,meV, highlighted by the red arrow, corresponds to a relatively flat dispersion in the vicinity of the M point. ARPES-measured (d) in-plane Fermi surface of (Sb\textsubscript{2}Te\textsubscript{3})\textsubscript{1}(Sb\textsubscript{2})\textsubscript{2}. 
    and electron energy dispersions along (e) $\overline{\Gamma}$ - $\overline{\mathrm{M}}$ and (f) $\overline{\mathrm{M}}$ - $\overline{\mathrm{K}}$ - $\overline{\Gamma}$. The bulk bands crossing the Fermi level near $\Gamma$ and M are labeled as G and M, with possible surface states labeled G' and M'. The corresponding sc$GW$-computed (g) in-plane Fermi surface and electron energy dispersions along (h) ${\Gamma}$ - ${\mathrm{M}}$ and (i) ${\mathrm{M}}$ - ${\mathrm{K}}$ - ${\Gamma}$, with bulk bands crossing the Fermi level near $\Gamma$ and M labeled as G and M.} 
    \label{fig:ARPES-GW}
\end{figure*}

We consider the computed band structure for antimony telluride homologous superlattices with 2 antimonene layers (Sb\textsubscript{2}Te\textsubscript{3})\textsubscript{1}(Sb\textsubscript{2})\textsubscript{2}. For the energy band dispersion in the vicinity of the $\Gamma$ point, shown in Fig.~\ref{fig:STS}(b), the band gap has closed. As will be discussed below, the computed band structure has been shifted upward by 0.6 eV to ensure alignment with key features at the $\Gamma$ and M-points in the valence band (Fig.~\ref{fig:ARPES-GW}(e)).  Thus, the band gap closing yields significant DOS at $\approx$ 0.6 eV above the Fermi level.  
 

We now discuss the origins of the additional states above the Fermi level. For this purpose, we consider the selected AO contributions to the computed DOS at three momenta within the vicinity of the band crossing: $\Gamma$ and $\pm$ 0.12 \AA\textsuperscript{-1} of $\Gamma$, i.e. at the dashed lines labeled $\Gamma$\textsubscript{+} and $\Gamma$\textsubscript{--} in Fig.~\ref{fig:STS}(b). For $\Gamma$\textsubscript{+} (Fig.~\ref{fig:STS}(c) top) and $\Gamma$\textsubscript{--} (Fig.~\ref{fig:STS}(c) bottom). 
Refer to the Supplemental Material for a more detailed decomposition of the DOS in AO.
The computed DOS at $\approx$ 0.2 eV is primarily associated with the hybridization of the antimonene Sb $p_z$ orbital with the antimony telluride layer. 
This supports the interpretation that interlayer $p_z$ orbital contribution drives the crossing of the sharp valance band with the original conduction band minimum.
It is unclear why the feature only diminishes in the n = 3 sample. 
The precise layer-number dependence of this feature likely reflects a more complex interplay of factors, which we leave for future study.

 
\section{Energy Band Dispersion}

For both antimony telluride and its homologous superlattices, the trigonal surface Brillouin zone (BZ) is shown in Fig.~\ref{fig:ARPES-GW}(a). With the addition of two layers of antimonene, antimony telluride homologous superlattices crystallize in the P-3m1 space group, resulting in the hexagonal bulk BZ (Fig.~\ref{fig:ARPES-GW}(a) middle). Meanwhile, for the antimony telluride homologous superlattices containing three or four layers of antimonene (R-3m space group), the rhombohedral bulk BZ (Fig.~\ref{fig:ARPES-GW}(a) bottom) is applicable. 

To examine the influence of antimonene on the energy band dispersion in antimony telluride homologous superlattices, we consider ARPES data in conjunction with sc$GW$ calculations. For antimony telluride homologous superlattices containing two layers of antimonene, (Sb\textsubscript{2}Te\textsubscript{3})\textsubscript{1}(Sb\textsubscript{2})\textsubscript{2}, the computed electron energy band dispersion along a path connecting the $\Gamma$, M, and K points is shown in Fig.~\ref{fig:ARPES-GW}(b). 
The computed band structure has been shifted upward by 0.6 eV to ensure alignment with key features at the $\Gamma$ and M-points in the bands observed by ARPES. 
Using the energy band dispersion in Fig.~\ref{fig:ARPES-GW}(b), we compute the density of states summed along the $\Gamma$ - M - K - $\Gamma$ path (Fig.~\ref{fig:ARPES-GW}(c)). 
The total DOS provides an approximate description of the d$I$/d$V$ signal. 
Strictly, d$I$/d$V$ is proportional to the local DOS evaluated at the tip position, which weights each state by its local real-space probability density.~\cite{tersoffTheoryApplicationScanning1983,tersoffTheoryScanningTunneling1985,krennerAssessmentScanningTunneling2013} 
Nevertheless, comparing the experimental d$I$/d$V$ spectrum for the n = 2 sample in Fig.~\ref{fig:STS}(a) with the computed total DOS in Fig.~\ref{fig:ARPES-GW}(c), we observe that the two curves share similar spectral profiles and features.
Interestingly, a sharp peak in the calculated DOS is observed at $\sim 200$\,meV below the $E\textsubscript{F}$, as highlighted by the red arrow.  
The peak in the DOS corresponds to a relatively flat dispersion in the vicinity of the M point (Fig.~\ref{fig:ARPES-GW}(b)). 
A van Hove singularity occurs when the group velocity vanishes at a critical point in the Brillouin zone ($\nabla_k E\approx 0$).~\cite{van1953occurrence}
The nearly zero slope near M in both the M-K and M-$\Gamma$ band path directions is consistent with the conditions for a van Hove singularity in 2D Brillouin zone. 

As shown in Fig.~\ref{fig:ARPES-GW}, the measured (Fig.~\ref{fig:ARPES-GW}(d)) and computed (Fig.~\ref{fig:ARPES-GW}(g))
Fermi surfaces of (Sb\textsubscript{2}Te\textsubscript{3})\textsubscript{1}(Sb\textsubscript{2})\textsubscript{2} exhibit the expected trigonal symmetry of the underlying crystal lattice, while containing multiple Fermi contours. Two hole pockets, including a circular contour centered around $\overline \Gamma$ and an oblong pocket oriented between $\overline \Gamma$ and $\overline{\mathrm{M}}$, are apparent in both the measured (Fig.~\ref{fig:ARPES-GW}(d)) and computed (Fig.~\ref{fig:ARPES-GW}(g)) Fermi surfaces. In addition, a comparison of the measured (Fig.~\ref{fig:ARPES-GW}(e)) and computed (Fig.~\ref{fig:ARPES-GW}(h)) band dispersions along $\Gamma$-M suggests that the G and M bands, which are observed in both cases, are derived from bulk bands, whereas the G' and M' bands, that are observed only in the measured dispersions, may be associated with surface states. 

We now consider the band dispersion in the vicinity of the M-point. For both the measured (Fig.~\ref{fig:ARPES-GW}(e)) and computed (Fig.~\ref{fig:ARPES-GW}(h)) band dispersions 
along $\overline \Gamma$ - $\overline{\mathrm{M}}$, positive curvature, with a local band mininum, is apparent near $\overline{\mathrm{M}}$. In contrast, for the perpendicular path ($\overline{\mathrm{K}}$ - $\overline{\mathrm{M}}$ - $\overline{\mathrm{K}}$) (Figs.~\ref{fig:ARPES-GW}(f,i)), negative curvature, with a local maximum, is apparent near $\overline{\mathrm{M}}$. The ARPES data yields inverted effective masses at $\overline{\mathrm{M}}$, with 1.7$m_e$ along $\overline \Gamma$ - $\overline{\mathrm{M}}$ and -0.1$m_e$ along $\overline{\mathrm{K}}$ - $\overline{\mathrm{M}}$ - $\overline{\mathrm{K}}$. Thus, the bands form a saddle point with energy $E\textsubscript{SP}=-255$\,meV (Fig.~\ref{fig:ARPES-GW}f).
The behavior is largely reproduced in our sc$GW$ calculations, although the calculated feature is flatter than the point-like feature observed in ARPES. Nonetheless, the sharp peak in the DOS slightly below $E\textsubscript{F}$ in Fig.~\ref{fig:ARPES-GW}(c) likely corresponds to the saddle point observed in ARPES.


\begin{figure}
    \centering
    \includegraphics[width=\linewidth]{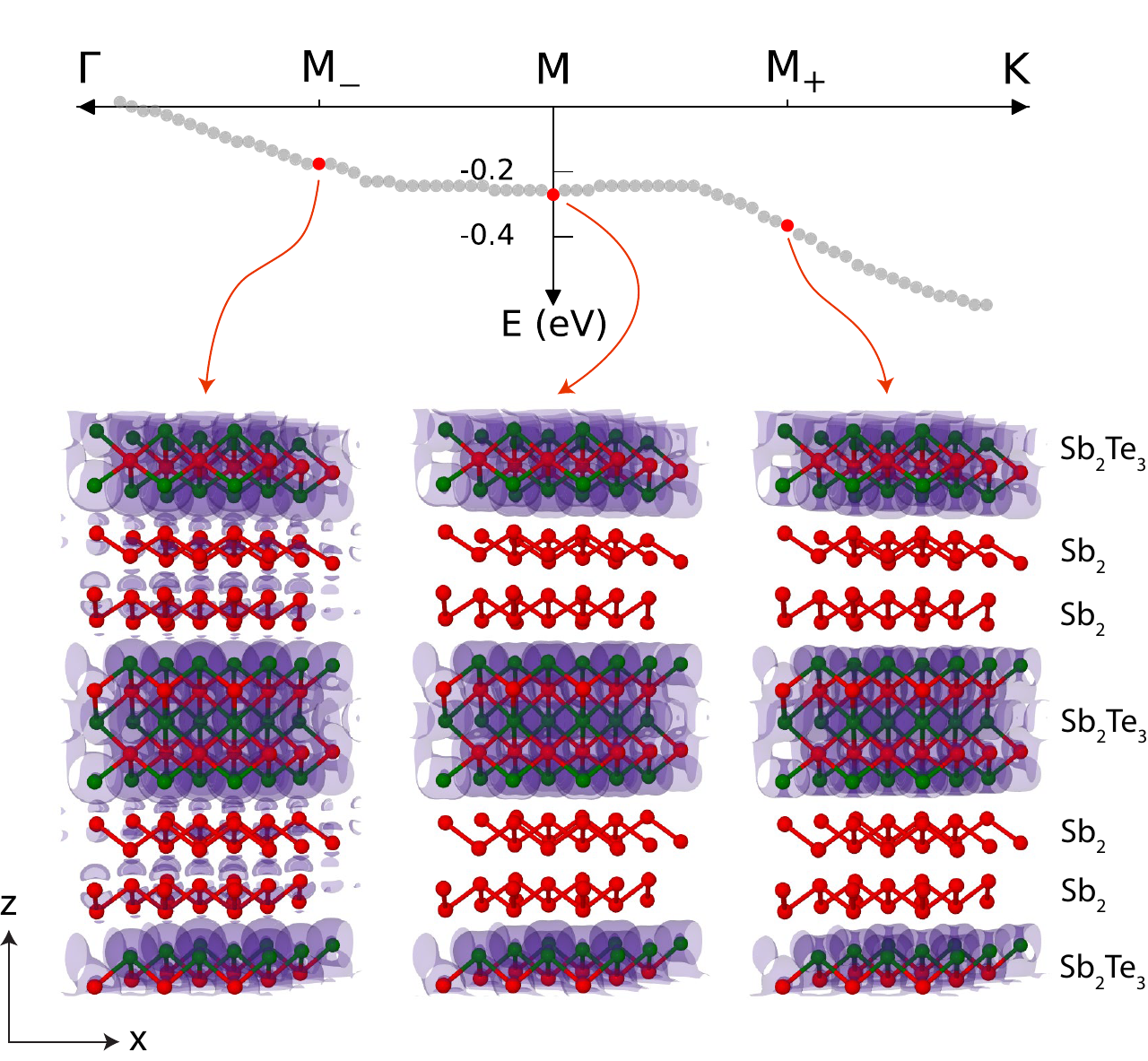}
    \caption{
    The band dispersion and electron density for (Sb\textsubscript{2}Te\textsubscript{3})\textsubscript{1}(Sb\textsubscript{2})\textsubscript{2} computed by sc$GW$ near the saddle point in the vicinity of M. (Top) Nearly flat band dispersion along the $\Gamma$-M-K path in the vicinity of the saddle point. Grey points correspond to maxima in the DOS, with red dots marking points within $\pm$ 0.23 \AA\textsuperscript{-1} of the M-point. 
    (Bottom) Real space depiction of the electron density at M\textsubscript{-}, M, and M\textsubscript{+}, constructed by summing all symmetrized atomic orbital contributions, each weighted by its spectral intensity fraction at the corresponding DOS peak.
    At M\textsubscript{-} and M, the electron density is primarily due to Te $p_x$ orbitals in the Sb$_{2}$Te$_{3}$ layer. At M\textsubscript{-}, the contributions of both Sb $p_z$ and Te $p_z$ orbitals are increased, and hybridization between Sb$_{2}$ and Sb$_{2}$Te$_{3}$ layers is apparent. 
    }
    \label{fig:AO orbital change with momentum}
\end{figure}

To understand the role of the antimonene layers in generating the saddle point, we consider the AO contributions to the computed DOS in the vicinity of $\overline{\mathrm{M}}$. The band dispersion along the $\Gamma$-M-K path in the vicinity of the saddle point is shown in Fig.~\ref{fig:AO orbital change with momentum}(top). In the plot, the grey symbols indicate maxima in the DOS, and the red dots mark points within $\pm$ 0.23 \AA\textsuperscript{-1} of the M-point, i.e. M\textsubscript{-}, M, and M\textsubscript{+}. The electron densities were constructed by summing all symmetrized atomic orbital contributions, each weighted by its spectral intensity fraction at the corresponding DOS peak, as described in the Supplemental Material. 

In Fig.~\ref{fig:AO orbital change with momentum}(bottom), the electron densities (purple) at M\textsubscript{-}, M, and M\textsubscript{+} are overlayed onto $3\times3\times2$ slabs of (Sb\textsubscript{2}Te\textsubscript{3})\textsubscript{1}(Sb\textsubscript{2})\textsubscript{2}. In the vicinity of the M-point, the electron density is dominated by in-plane Te $p_x$ orbitals within the Sb$_2$Te$_3$ layer. 
The saddle point emerges along the K--M--$\Gamma$ path as a consequence of enhanced interlayer $p_z$ hybridization between the Sb$_2$ and Sb$_2$Te$_3$ layers, which induces opposite band curvatures along orthogonal momentum directions.
Although the antimonene layers generate increased interlayer $p_z$ hybridization that moves the band towards the Fermi level, the overall band dispersion remains largely unchanged because the electron density character is still dominated by the Sb$_2$Te$_3$ layer.

\begin{figure*}[ht]
    \centering
    \includegraphics[width=\linewidth]{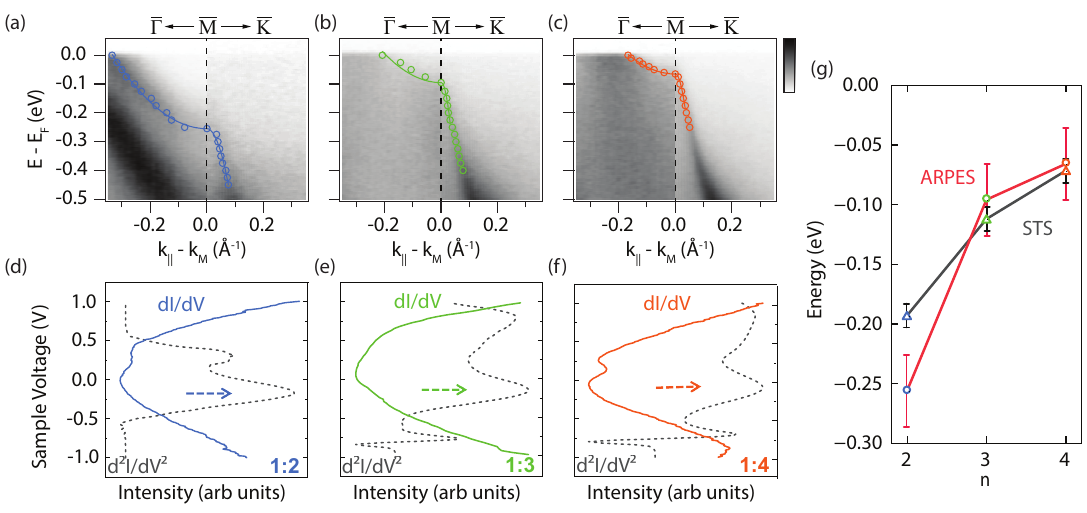}
    \caption{Evolution of saddle point in antimony telluride homologous superlattices containing 2, 3, and 4 antimonene layers, i.e. (Sb$_{2}$Te$_{3}$)$_{1}$(Sb$_{2}$)$_{n}$ (n = 2, 3, 4): (a)-(c) ARPES-measured electron energy dispersions along the $\overline{\mathrm{\Gamma}}$ - $\overline{\mathrm{M}}$ - $\overline{\mathrm{K}}$ path in the vicinity of $\overline{\mathrm{M}}$ for antimonene telluride homologous superlattices with (a) 2, (b) 3, and (c) 4 antimonene layers. The extracted band positions (open circles) and their parabolic fits (solid lines) are overlaid for each plot. The color bar indicates ARPES intensity from minimum (bottom) to maximum (top). (d)-(e) STS bias voltage vs. differential conductance (d$I$/d$V$, colored curves) and second-harmonic (d$^2I$/d$V^2$, black dotted curves) for antimony telluride homologous superlattices with (d) 2, (e), 3, and (f) 4 antimonene layers. The arrows indicate peaks in the second-harmonic spectra, near the Fermi level, corresponding to the van Hove singularities. (g) Saddle-point energies extracted from ARPES and the corresponding evolution of the van Hove singularity energy determined from the d$^2I$/d$V^2$ spectra as a function of $n$ in (Sb$_{2}$Te$_{3}$)$_{1}$(Sb$_{2}$)$_{n}$.} 
    \label{fig:evolution}
\end{figure*}

\section{Evolution of saddle point}

We now examine the influence of increasing number of antimonene layers on the evolution of the electronic band structure and the saddle point at $\overline{\mathrm{M}}$. ARPES data reveal similar electron energy band dispersions across the antimony telluride homologous superlattice series with 2 to 4 antimonene layers, i.e. (Sb\textsubscript{2})\textsubscript{n} ($n = 2, 3, 4$), including two hole pockets derived from the bulk G and M bands (Supplementary Material) and the saddle point at $\overline{\mathrm{M}}$. In Figs.~\ref{fig:evolution}(a)-(c), we present the energy band dispersion along the $\overline{\mathrm{\Gamma}}$ - $\overline{\mathrm{M}}$ - $\overline{\mathrm{K}}$ path in the vicinity of $\overline{\mathrm{M}}$, with the blue, green, and orange circles highlighting the saddle point for (Sb\textsubscript{2})\textsubscript{n} with n = 2, 3, and 4.  Interestingly, the saddle point energy and its dispersion evolves across the homologous superlattice series. Notably, the energy of the saddle point $E\textsubscript{SP}$ shifts monotonically towards $E\textsubscript{F}$ as a function of $n$ from $E_{SP}\,=\,-255$\,meV ($n=2$) to $E_{SP}\,=\,-65$\,meV ($n\,=\,4$) (Fig.~\ref{fig:evolution} g). For $\overline{\mathrm{M}}$ - $\overline \Gamma$ - $\overline{\mathrm{M}}$, the effective mass evolves from 1.7$m_e$ (n=2) to 1.5$m_e$ (n=3)  to 1.6$m_e$ (n=4).  Meanwhile, for $\overline{\mathrm{K}}$ - $\overline{\mathrm{M}}$ - $\overline{\mathrm{K}}$, the effective mass evolves from  -0.1$m_e$ (n=2) to -0.03$m_e$ (n=3) to -0.04$m_e$ (n=4).  Thus, there is an effective mass asymmetry between the $\overline \Gamma$ - $\overline{\mathrm{M}}$ and $\overline{\mathrm{M}}$ - $\overline{\mathrm{K}}$ - $\overline{\mathrm{M}}$ paths across the series. 
  

To probe van Hove singularities in the density of states associated with the observed saddle points, we utilize the second-harmonic of the tunneling current, (d$^2I$/d$V^2$), which provides enhanced sensitivity to subtle changes in the energy dependence of the local density of states.  In Figs.~\ref{fig:evolution} (d-f), the STS sample bias voltage is plotted as a function of both the differential conductance (d$I$/d$V$, colored curves) and the second-harmonic (d$^2I$/d$V^2$, black dotted curves) for antimony telluride homologous superlattices with n = (d) 2, (e) 3, and (f) 4.  The sample voltages correspond to the energy relative to the Fermi level. 


For the d$^2I$/d$V^2$ spectra, blue, green, and orange arrows indicate distinct features at energies that coincide with those of the saddle points in the electronic dispersion identified in ARPES.  Since saddle points in the electronic dispersion give rise to van Hove singularities in the density of states, these maxima in d$^2I$/d$V^2$ reflect rapid variation and enhancement of the local density of states associated with van Hove singularity.  Furthermore, as n increases, the energy of the van Hove singularity systematically shifts toward the Fermi level (E\textsubscript{F}), similar to the ARPES-determined trend for saddle point energies (Fig.~\ref{fig:evolution}(g)).

\section{Summary and Outlook}
In summary, we examined the effects of antimonene layers on the electronic structure of antimony telluride homologous superlattices. Using single crystals prepared by solution growth and solid state synthesis, we identified the crystal structures based upon long-period stacking sequences quantified via powder x-ray diffraction.
To isolate the effects of antimonene layers on the semiconductor to semi-metal transition, we consider the differential conductance for antimony telluride homologous superlattices in comparison with that of antimony telluride. With the addition of antimonene layers, the energy band gap closes and additional states emerge above the Fermi level. Self-consistent $GW$ calculations suggest that the antimonene-generated bandgap closing is driven by interlayer hybridization of Sb $p_z$ orbitals at the $\Gamma$ point.  

A comparison of ARPES-measured and sc$GW$-computed Fermi surfaces reveal multiple Fermi contours including two hole pockets, along with a saddle point at M. 
As the number of antimonene layers is increased from two to four, the saddle-point energies extracted from ARPES and the corresponding energies of the van Hove singularities determined from the second-harmonic STS spectra (i.e. d$^2I$/d$V^2$) move toward the Fermi level.
We identify the key role of Sb and Te $p_z$ orbital hybridization in generating the saddle point and shifting it to higher energies with increasing number of antimonene layers.  

More broadly, we demonstrate the ability to systematically bring a van Hove singularity into near alignment with the Fermi level using homologous superlattices consisting of topological insulators and topological semimetals.  This homologous superlattice approach provides a promising platform for exploring correlated quantum phenomena without the need for gating or doping. As the density of states near the Fermi level is enhanced, the homologous superlattices may host emergent electronic instabilities such as unconventional superconductivity, magnetism, or charge density waves. Furthermore, the homologous superlattice approach is applicable to a wide range of systems, especially those combining topological insulators with topological semimetals. 


\section*{Acknowledgments}
This research was supported by the National Science Foundation (NSF) through the Materials Research Science and Engineering Center at the University of Michigan, Award No. DMR-2309029. This material is also based upon work supported by the NSF CAREER grant under Award No. DMR-2337535. 

This work used resources of the Advanced Light Source, a U.S. Department of Energy (DOE) Office of Science User Facility under Contract No. DE-AC02-05CH11231. 

The authors gratefully acknowledge Robert Hovden and Nishkarsh Agarwal for insightful discussions.

The authors gratefully acknowledge Tao-Yu Huang and Gray Schneider for development of the project.

\section*{Conflict of interest}
The authors have no conflicts of interest to disclose.

\section*{References}

\end{document}


\title{Supporting Information: Evolution of the Saddle Point in Antimony Telluride Homologous Superlattices}

\author{Yi-Hsin Shen}
\affiliation{Applied Physics Program, University of Michigan, Ann Arbor, MI 48109, USA}

\author{Shane Smolenski}
\affiliation{Department of Physics, University of Michigan, Ann Arbor, MI 48109, USA}

\author{Ming Wen}
\affiliation{Department of Chemistry, University of Michigan, Ann Arbor, MI 48109, USA}

\author{Yimo Hou}
\affiliation{Department of Materials Science and Engineering, University of Michigan, Ann Arbor, MI 48109, USA}

\author{Eoghan Downey}
\affiliation{Department of Physics, University of Michigan, Ann Arbor, MI 48109, USA}

\author{Jakob Hammond-Renfro}
\affiliation{Department of Physics, University of Michigan, Ann Arbor, MI 48109, USA}

\author{Katharine Moncrieffe}
\affiliation{Department of Electrical Engineering, Cooper Union for the Advancement of Science and Art, New York City, NY 10003, USA}

\author{Chun Lin}
\affiliation{Stanford Synchrotron Radiation Lightsource, SLAC National Accelerator Laboratory, Menlo Park, CA 94025, USA}

\author{Makoto Hashimoto}
\affiliation{Stanford Synchrotron Radiation Lightsource, SLAC National Accelerator Laboratory, Menlo Park, CA 94025, USA}

\author{Donghui Lu}
\affiliation{Stanford Synchrotron Radiation Lightsource, SLAC National Accelerator Laboratory, Menlo Park, CA 94025, USA}

\author{Kai Sun}
\affiliation{Department of Physics, University of Michigan, Ann Arbor, MI 48109, USA}

\author{Dominika Zgid}
\affiliation{Department of Physics, University of Michigan, Ann Arbor, MI 48109, USA}
\affiliation{Department of Chemistry, University of Michigan, Ann Arbor, MI 48109, USA}
\affiliation{Institute of Theoretical Physics, University of Warsaw, 02-093 Warsaw, Poland }

\author{Emanuel Gull}
\affiliation{Department of Physics, University of Michigan, Ann Arbor, MI 48109, USA}
\affiliation{Institute of Theoretical Physics, University of Warsaw, 02-093 Warsaw, Poland }

\author{Pierre Ferdinand P. Poudeu}
\affiliation{Department of Materials Science and Engineering, University of Michigan, Ann Arbor, MI 48109, USA}

\author{Na Hyun Jo}
\altaffiliation{nhjo@umich.edu}
\affiliation{Department of Physics, University of Michigan, Ann Arbor, MI 48109, USA}

\author{Rachel S. Goldman}
\altaffiliation{rsgold@umich.edu }
\affiliation{Department of Materials Science and Engineering, University of Michigan, Ann Arbor, MI 48109, USA}
\affiliation{Department of Physics, University of Michigan, Ann Arbor, MI 48109, USA}

\maketitle

\setcounter{equation}{0}
\setcounter{figure}{0}
\setcounter{table}{0}
\setcounter{page}{1}
\makeatletter
\renewcommand{\theequation}{S\arabic{equation}}
\renewcommand{\thefigure}{S\arabic{figure}}
\renewcommand{\thetable}{S\arabic{table}}

\newcommand{\postit}[1]{
    \begin{center}
        \fbox{
            \begin{minipage}{0.8\textwidth}
                \textit{#1}
            \end{minipage}
            }
    \end{center}
}

\clearpage

\tableofcontents

\clearpage

\section{Additional computation detail}

\subsection{Decomposition of DOS}

The self-consistent $GW$ (sc$GW$) calculation was performed with Gaussian type orbital (GTO) basis. 
The density of states (DOS) was transformed to the symmetrized atomic orbital (SAO) basis to plot the band structure in FIG.~4 of the main text. 

\begin{figure}[hb]
    \centering
    \includegraphics[width=1.0\linewidth]{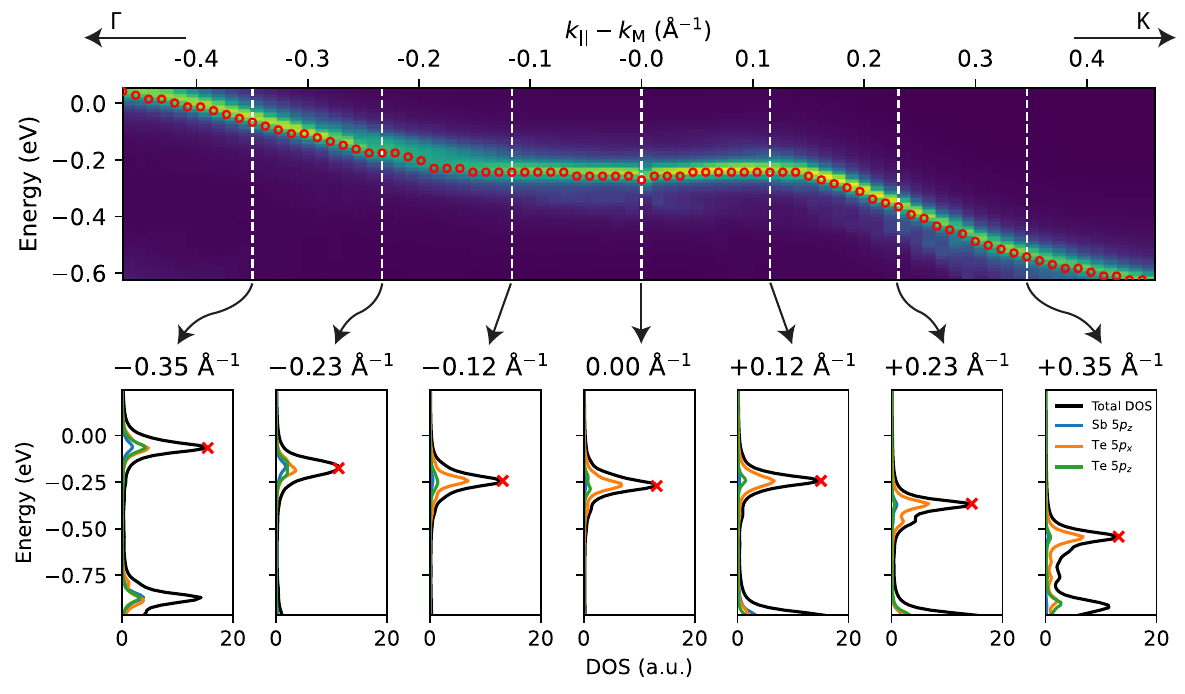}
    \caption{
    (Top) The band dispersion near high symmetry point M calculated with sc$GW$. Red circles denoted peaks. 
    (Bottom) The DOS composition at different points along the band. Sb 5$p_z$ and Te 5$p_x$ and 5$p_z$ contributions are plotted. 
    Red crosses denoted the same peak.
    The electron densities in Fig.~4 of the main text corresponds to the DOS at $k_{||} - k_{\mathrm{M}} = 0$ (denoted as M) and $\pm 0.23$ \AA$^{-1}$ (denoted as M$_+$ and M$_-$). }
    \label{fig:all_dos}
\end{figure}

\begin{figure}
    \centering
    \includegraphics[width=0.7\linewidth]{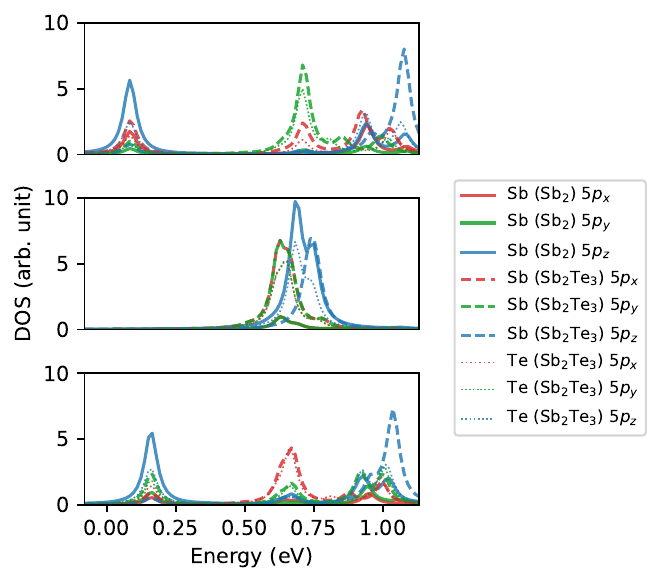}
    \caption{
    DOS contribution from $p$ atomic orbitals calculated with sc$GW$ at $\Gamma$\textsubscript{+} (top), $\Gamma$ (middle), and $\Gamma$\textsubscript{--} (bottom). The same plot is presented as FIG.~2(c) of the main text with only selected Sb $p$ orbitals for clarity.}
    \label{fig:all_dos_band_closing}
\end{figure}

Atomic orbital contributions to the saddle point were calculated as follows: 
We first found the peak height percentages $p^\mathrm{SAO}(E)$ at the high symmetry point M and points near it. 
At any point $(\vec k,E)$ on band structure plot, the total peak heights can be attributed to all atomic orbitals (AO) via the normalized coefficient vectors $C^\mathrm{AO}$ as
\begin{equation}
    p_{i}^\mathrm{AO} = \sum_\alpha p_{\alpha}^\mathrm{SAO} [C_{\alpha i}^\mathrm{AO}]^2,
\end{equation}
in which $i$ and $\alpha$ are indices for AO and SAO respectively. 
The AO percentages at a fixed ($\vec k,E$) are normalized as
\begin{equation}
    \sum_ip_{i}^\mathrm{AO} = 1.
\end{equation}
In order to visualize the electron density at $(\vec k,E)$, we construct the full weighted density matrix as
\begin{equation}
    \rho = [C^{\mathrm{AO}}]^{\mathrm{T}}\mathrm{diag}(p^{\mathrm{SAO}})C^{\mathrm{AO}}
\end{equation}
Using the internal functions in \texttt{pyscf},~\cite{sunLibcintEfficientGeneral2015b,sunPySCFPythonbasedSimulations2018a,sunRecentDevelopmentsPySCF2020a} the calculated density matrix is loaded into \texttt{cube} files, which were used to plot the electron density in FIG.~\ref{fig:all_dos}.

In FIG.~\ref{fig:all_dos_band_closing}, the AO contributions are resolved into Te $p$ and Sb $p$ components respectively. 
The band structure at high symmetry point $\Gamma$ has a dominant contribution from Sb $p_z$ orbital both from the Sb$_2$Te$_3$ and Sb$_2$ layers. 
There is a small but finite band gap in the pristine Sb$_2$Te$_3$ (see the $n = 0$ curve of FIG.~2(a) in the main text).
By adding Sb$_2$ layers to Sb$_2$Te$_3$, the small band gap closes, driven primarily by interlayer hybridization of the $p_z$ orbitals.

\subsection{2-D DOS iso-energy surfaces}

\begin{figure}[hb]
    \centering
    \includegraphics[width=0.8\linewidth]{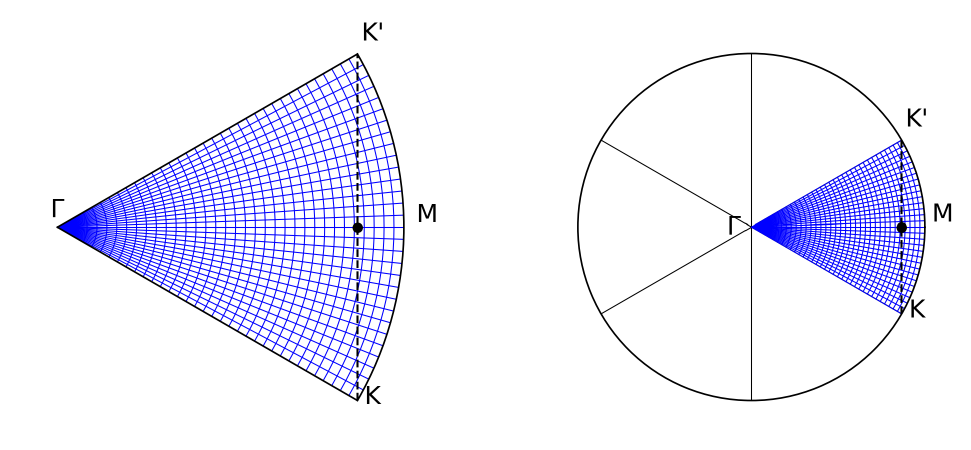}
    \caption{(Left) The sector-shaped area in the reciprocal space spanning K$-\Gamma-$K'. 
    (Right) The K$-\Gamma-$K' sector area is repeated five times by rotating it by 60, 120, 180, 240, and 300 degrees.}
    \label{fig:sector_DOS}
\end{figure}

\begin{figure}[ht]
    \centering
    \includegraphics[width=1.0\linewidth]{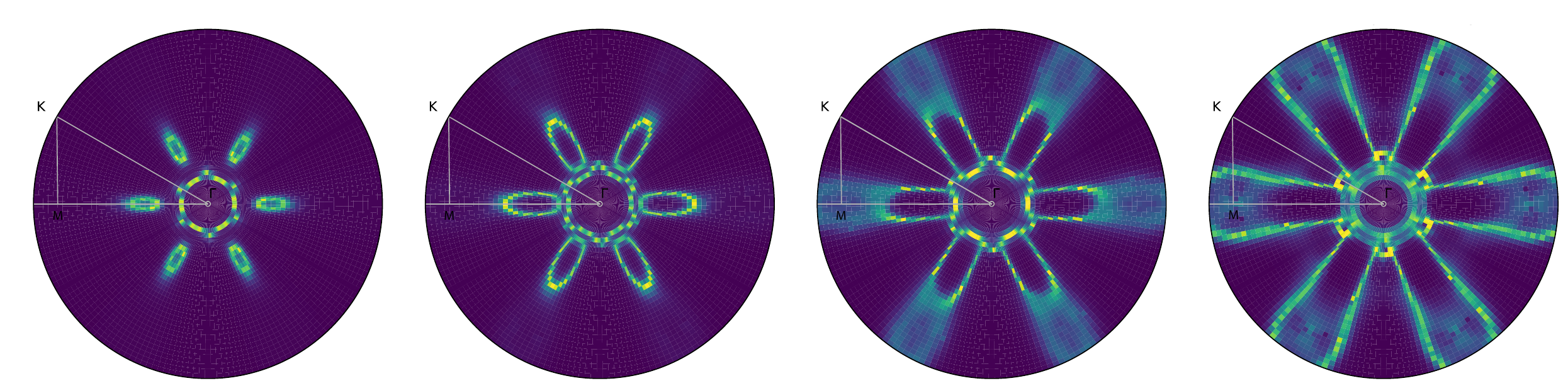}
    \caption{From left to right, each 2D DOS iso-surface plot represents the energy at $E_\mathrm{F}$, $E_\mathrm{F} - 0.1$ eV, $E_\mathrm{F} - 0.2$ eV, $E_\mathrm{F} - 0.3$ eV. calculated by sc$GW$.}
    \label{fig:fermi_0_to_0.3}
\end{figure}

We use the similar technique reported in Ref.~\citenum{smolenskiLargeExcitonBinding2025a} to render the 2-D DOS iso-energy planes. 
A $36 \times 31$ $k$-mesh grid in a sector-shaped area spanning K$-\Gamma-$K' is built as in FIG.~\ref{fig:sector_DOS}.
The Nevanlinna analytic continuation is performed for every vector starting from $\Gamma$ for every 2 degrees. 
The full 2-D DOS iso-energy surface in FIG.~4(g) is plotted by rotating the K$-\Gamma-$K' DOS five times following the hexagonal lattice symmetry. 
In FIG.~\ref{fig:fermi_0_to_0.3}, we showcase some additional 2-D DOS iso-energy surface from $E_\mathrm{F}$ to $E_F-0.3$ eV.

\clearpage

\section{Additional ARPES detail}

\subsection{Surface Termination Identification}

\begin{figure}[ht]
    \centering
    \includegraphics[width=0.75\linewidth]{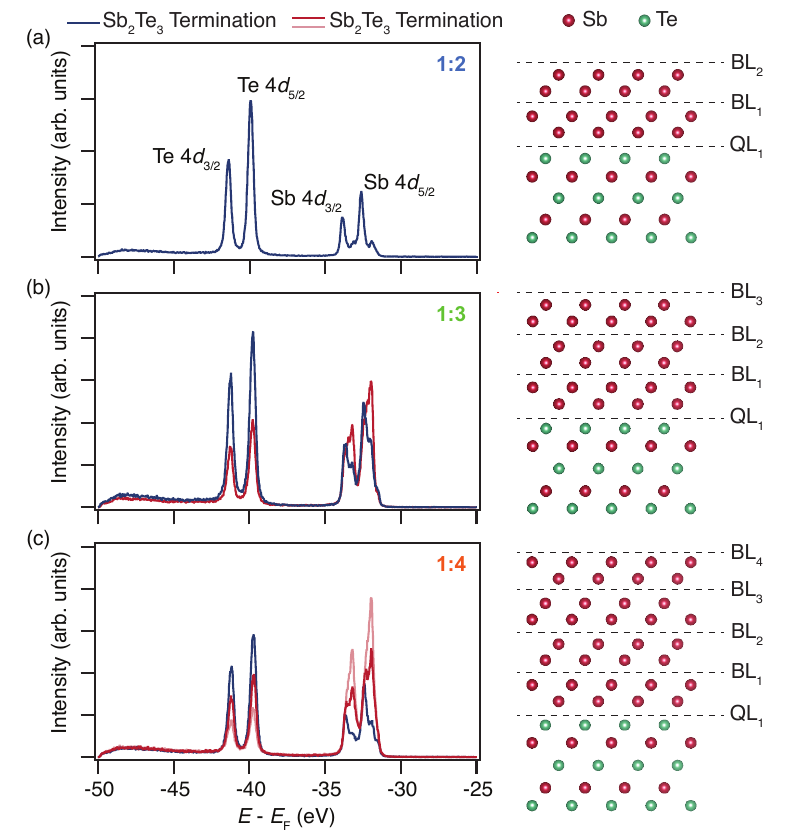}
    \caption{Representative photoemission spectra around the Te 4\textit{d} and Sb 4\textit{d} core levels found on the cleaved surfaces of (a) (Sb\textsubscript{2}Te\textsubscript{3})\textsubscript{1}(Sb\textsubscript{2})\textsubscript{2}, (b)(Sb\textsubscript{2}Te\textsubscript{3})\textsubscript{1}(Sb\textsubscript{2})\textsubscript{3}, and (c) (Sb\textsubscript{2}Te\textsubscript{3})\textsubscript{1}(Sb\textsubscript{2})\textsubscript{4}. Schematics illustrating the different possible surface terminations for each compound are shown at right. 
    } 
    \label{fig:XPS}
\end{figure}

Due to the surface sensitivity of ARPES and the complex crystal structure of (Sb\textsubscript{2}Te\textsubscript{3})\textsubscript{m}(Sb\textsubscript{2})\textsubscript{n}, it is important to measure a consistent surface termination for each sample to properly compare across the series. For (Sb\textsubscript{2}Te\textsubscript{3})\textsubscript{m}(Sb\textsubscript{2})\textsubscript{n}, the structure consists of layers of Sb\textsubscript{2}Te\textsubscript{3} quintuple layers (QLs) and Sb\textsubscript{2} bilayers (BLs). As the bonds within QLs and BLs are covalent in nature whereas the bonds between QLs and BLs and between two BLs are van der Waals~\cite{johannsen2015engineering,GOVAERTS2012,LIND2005}, the expected cleavage planes are between neighboring QLs and BLs. Thus, the expected surface terminations are relatively Te-rich QLs and relatively Sb-rich BLs. For (Sb\textsubscript{2}Te\textsubscript{3})\textsubscript{1}(Sb\textsubscript{2})\textsubscript{n}, a single QL termination exists while there are n different BL terminations. The likely surface terminations for each compound in the (Sb\textsubscript{2}Te\textsubscript{3})\textsubscript{1}(Sb\textsubscript{2})\textsubscript{n} series for $n=2-4$ are shown in Fig.~\ref{fig:XPS}.

The surface termination can be determined by comparing the relative contributions of Sb 4\textit{d} and Te 4\textit{d} core levels~\cite{johannsen2015engineering}. More specifically, QL terminations will exhibit relatively larger Te 4\textit{d} photoemission intensity than BL terminations. In Fig.~\ref{fig:XPS}, we present representative photoemission spectra around the Sb 4\textit{d} and Te 4\textit{d} core levels from different regions on the cleaved surfaces of (Sb\textsubscript{2}Te\textsubscript{3})\textsubscript{1}(Sb\textsubscript{2})\textsubscript{2}, (Sb\textsubscript{2}Te\textsubscript{3})\textsubscript{1}(Sb\textsubscript{2})\textsubscript{3}, and (Sb\textsubscript{2}Te\textsubscript{3})\textsubscript{1}(Sb\textsubscript{2})\textsubscript{4}. For (Sb\textsubscript{2}Te\textsubscript{3})\textsubscript{1}(Sb\textsubscript{2})\textsubscript{2}, we observe a single representative spectrum across the cleaved surface, indicating a uniform surface termination. For (Sb\textsubscript{2}Te\textsubscript{3})\textsubscript{1}(Sb\textsubscript{2})\textsubscript{3} and (Sb\textsubscript{2}Te\textsubscript{3})\textsubscript{1}(Sb\textsubscript{2})\textsubscript{4}, we observe multiple representative spectra, indicative of different surface terminations. Spectra with dominant intensity arising from Te 4\textit{d} core levels (blue curves) are ascribed to the QL termination while those with dominant Sb 4\textit{d} contributions are ascribed to BL terminations (red curves)~\cite{johannsen2015engineering}. While we only observe Sb 4\textit{d}-dominant regions with a single Sb 4\textit{d}:Te 4\textit{d} intensity ratio in (Sb\textsubscript{2}Te\textsubscript{3})\textsubscript{1}(Sb\textsubscript{2})\textsubscript{3}, we observe two representative ratios in (Sb\textsubscript{2}Te\textsubscript{3})\textsubscript{1}(Sb\textsubscript{2})\textsubscript{4}, possibly indicating different BL terminations.  

Having identified these distinct regions on each sample, all ARPES measurements were performed on regions identified as having a QL surface termination to maintain consistency for measurements across the (Sb\textsubscript{2}Te\textsubscript{3})\textsubscript{1}(Sb\textsubscript{2})\textsubscript{n} series.

\subsection{Photon energy dependence}

\begin{figure}[ht]
    \centering
    \includegraphics[width=\linewidth]{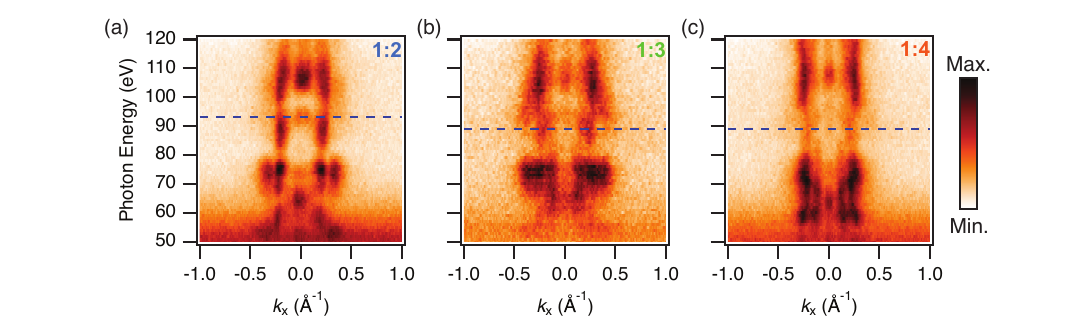}
    \caption{Photon energy dependence of (Sb\textsubscript{2}Te\textsubscript{3})\textsubscript{1}(Sb\textsubscript{2})\textsubscript{3} and (Sb\textsubscript{2}Te\textsubscript{3})\textsubscript{1}(Sb\textsubscript{2})\textsubscript{4}. (a, b, c) Iso-energy contours in the $k_x$ - $h\nu$ plane taken at an energy near the bottom of the G' band for (a) (Sb\textsubscript{2}Te\textsubscript{3})\textsubscript{1}(Sb\textsubscript{2})\textsubscript{2}, (b) (Sb\textsubscript{2}Te\textsubscript{3})\textsubscript{1}(Sb\textsubscript{2})\textsubscript{3}, and (c) (Sb\textsubscript{2}Te\textsubscript{3})\textsubscript{1}(Sb\textsubscript{2})\textsubscript{4}. The dashed lines indicate the photon energies used to measure the in-plane dispersion for each compound. 
    } 
    \label{fig:ARPES_hv}
\end{figure}

We performed photon energy-dependent measurements on (Sb\textsubscript{2}Te\textsubscript{3})\textsubscript{1}(Sb\textsubscript{2})\textsubscript{n} ($n = 2, 3, 4$) in order to probe the energy dispersion along $k_z$ and to optimize photoemission intensity. The iso-energy contours in the $k_x$ - $h\nu$ plane (where $h\nu$ is directly related to $k_z$ in the final state free electron approximation) taken at an energy near the bottom of the G’ band for each compound are presented in Fig.~\ref{fig:ARPES_hv}. The contours for each compound, which each exhibit nearly straight features with strong intensity modulations with photon energy, primarily consist of the G’ (features near $k_x = 0\text{\AA}^{-1}$) and G (features near $k_x = 0.25\text{\AA}^{-1}$)) bands. While the periodicity of the intensity modulations for (Sb\textsubscript{2}Te\textsubscript{3})\textsubscript{1}(Sb\textsubscript{2})\textsubscript{2} appear to be in moderate agreement with the expected $k_z$ periodicity corresponding to the theoretical \textit{c} lattice parameter of $17.633\text{\AA}$, no clear modulations commensurate with the lattice parameters of $63.896\text{\AA}$ and $75.51\text{\AA}$ are observed in (Sb\textsubscript{2}Te\textsubscript{3})\textsubscript{1}(Sb\textsubscript{2})\textsubscript{3} or (Sb\textsubscript{2}Te\textsubscript{3})\textsubscript{1}(Sb\textsubscript{2})\textsubscript{4}, respectively. Additionally, while intensity distribution of the bands changes significantly with photon energy, many bands are observed across wide ranges of photon with minimal changes in energy for all compounds, likely due to minimal $k_z$ dispersion related to the van der Waals nature of the compounds and/or significant $k_z$ broadening given the photon energies used and large \textit{c} lattice parameters. Given these observations, we selected photon energies that maximized the intensity of the bands of interest, namely the M band forming the saddle point near $\overline{\mathrm{M}}$.

\subsection{Electronic structure of (Sb\textsubscript{2}Te\textsubscript{3})\textsubscript{1}(Sb\textsubscript{2})\textsubscript{3} and (Sb\textsubscript{2}Te\textsubscript{3})\textsubscript{1}(Sb\textsubscript{2})\textsubscript{4}}

\begin{figure}[ht]
    \centering
    \includegraphics[width=0.5\linewidth]{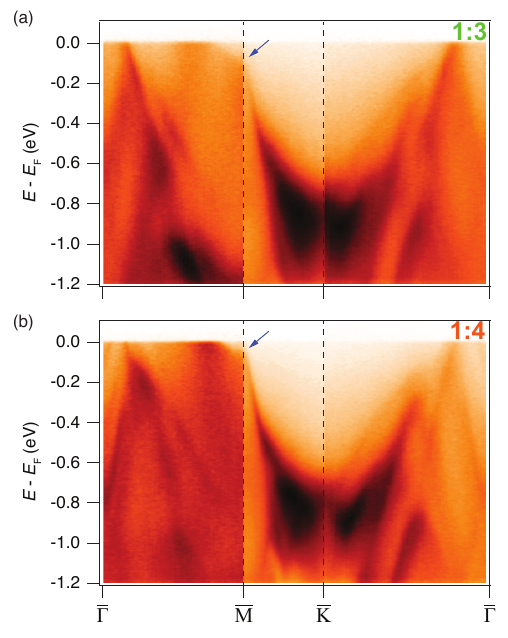}
    \caption{Electronic structure of (Sb\textsubscript{2}Te\textsubscript{3})\textsubscript{1}(Sb\textsubscript{2})\textsubscript{3} and (Sb\textsubscript{2}Te\textsubscript{3})\textsubscript{1}(Sb\textsubscript{2})\textsubscript{4}. (a, b) Measured electronic energy dispersion along select high symmetry paths Bulk and surface Brillouin zones for (a) (Sb\textsubscript{2}Te\textsubscript{3})\textsubscript{1}(Sb\textsubscript{2})\textsubscript{3} and (b) (Sb\textsubscript{2}Te\textsubscript{3})\textsubscript{1}(Sb\textsubscript{2})\textsubscript{4}.
    } 
    \label{fig:ARPES_13and14}
\end{figure}

We present the electronic band dispersion along high-symmetry paths for (Sb\textsubscript{2}Te\textsubscript{3})\textsubscript{1}(Sb\textsubscript{2})\textsubscript{3} and (Sb\textsubscript{2}Te\textsubscript{3})\textsubscript{1}(Sb\textsubscript{2})\textsubscript{4} in Fig.~\ref{fig:ARPES_13and14}. The dominant bulk bands are qualitatively similar to that of (Sb\textsubscript{2}Te\textsubscript{3})\textsubscript{1}(Sb\textsubscript{2})\textsubscript{2}. More specifically, the Fermi surface consists of two hole pockets derived from the bulk G and M bands in both (Sb\textsubscript{2}Te\textsubscript{3})\textsubscript{1}(Sb\textsubscript{2})\textsubscript{3} and (Sb\textsubscript{2}Te\textsubscript{3})\textsubscript{1}(Sb\textsubscript{2})\textsubscript{4}, similar to (Sb\textsubscript{2}Te\textsubscript{3})\textsubscript{1}(Sb\textsubscript{2})\textsubscript{2}. The surface G’ band is also faintly present in both compounds although we do not observe the M’ band in either (Sb\textsubscript{2}Te\textsubscript{3})\textsubscript{1}(Sb\textsubscript{2})\textsubscript{3} or (Sb\textsubscript{2}Te\textsubscript{3})\textsubscript{1}(Sb\textsubscript{2})\textsubscript{4}. Importantly, a saddle point is also observed at $\overline{\mathrm{M}}$ in both (Sb\textsubscript{2}Te\textsubscript{3})\textsubscript{1}(Sb\textsubscript{2})\textsubscript{3} and (Sb\textsubscript{2}Te\textsubscript{3})\textsubscript{1}(Sb\textsubscript{2})\textsubscript{4}. While the dominant electronic features are qualitatively similar across the (Sb\textsubscript{2}Te\textsubscript{3})\textsubscript{1}(Sb\textsubscript{2})\textsubscript{n} series for $n=2-4$, the band energies, including the saddle point, are different for each compound, as described in the Main Text.

\clearpage

\providecommand{\latin}[1]{#1}
\makeatletter
\providecommand{\doi}
  {\begingroup\let\do\@makeother\dospecials
  \catcode`\{=1 \catcode`\}=2 \doi@aux}
\providecommand{\doi@aux}[1]{\endgroup\texttt{#1}}
\makeatother
\providecommand*\mcitethebibliography{\thebibliography}
\csname @ifundefined\endcsname{endmcitethebibliography}  {\let\endmcitethebibliography\endthebibliography}{}